\documentclass[10pt,a4paper]{article}
\usepackage[utf8]{inputenc}
\usepackage[english]{babel}

\usepackage{amsmath}
\usepackage{amsfonts}
\usepackage{amssymb}

\usepackage[colorlinks,citecolor=blue,urlcolor=blue,linkcolor=blue]{hyperref}

\usepackage[left=2cm,right=2cm,top=2cm,bottom=2cm]{geometry}

\usepackage{authblk}

\usepackage{feynmp}
\newcommand{\pd}{\partial}

\title{Beyond Horndeski interactions induced by quantum effects}
\author[1,2]{B. Latosh \thanks{latosh@theor.jinr.ru}}
\affil[1]{Bogoliubov Laboratory of Theoretical Physics, JINR, Dubna 141980, Russia}
\affil[2]{Dubna State University, Universitetskaya str. 19, Dubna 141982, Russia}
\date{}

\begin{document}

\maketitle

\begin{abstract}
  Opportunity to generate beyond Horndeski interactions is addressed. An amplitude generating a certain beyond Horndeski coupling is explicitly found. The amplitude is free from ultraviolet divergences, so it is protected from ultraviolet contributions and can be considered as a universal prediction of effective field theory.
\end{abstract}

\section{Introduction}

Horndeski models are scalar-tensor models that have second-order field equations and have no interaction between the scalar field and matter \cite{Horndeski:1974wa,Kobayashi:2011nu}. Because of the second-order field equations they are protected from the Ostrogradsky instability associated with higher derivatives \cite{Ostrogradsky:1850fid}. The Horndeski action reads (Jordan frame):
\begin{align}\label{Horndeski_action}
  \mathcal{A} =& \int d^4 x \sqrt{-g} \left[\mathcal{L}_2 +\mathcal{L}_3+\mathcal{L}_4 + \mathcal{L}_5 +\mathcal{L}_\text{matter}[\Psi, g_{\mu\nu}]\right] ,
\end{align}
\begin{align*}
  \mathcal{L}_2 =& G_2 , \\
  \mathcal{L}_3 =& G_3\, \square \phi, \\
  \mathcal{L}_4 =& G_4\,R + G_{4,X}\, [(\square\phi)^2 - (\nabla_{\mu\nu}\phi)^2], \\
  \mathcal{L}_5 =& G_5\, G^{\mu\nu} \nabla_{\mu\nu}\phi - \cfrac16\,G_{5,X}\, \left[(\square\phi)^3 - 3 \, (\nabla_{\mu\nu}\phi)^2 \, \square \phi + 2 (\nabla_{\mu\nu}\phi)^3\right] .
\end{align*}
Here $\Psi$ notes matter degrees of freedom; $G_i$ are arbitrary functions of the scalar field $\phi$ and its canonical kinetic term $X=1/2\, (\pd\phi)^2$; $G_{i,X}$ are derivatives with respect to $X$; $R$ is the Ricci scalar, and $G_{\mu\nu}$ is the Einstein tensor. It should be noted these Lagrangians are extensively studied and they are constraint with empirical data \cite{Ezquiaga:2017ekz,TheLIGOScientific:2017qsa,Copeland:2018yuh,Latosh:2020jyq}. We will not discuss these constraints in details as they lie beyond the scope of this paper.

Beyond Horndeski models generalize Horndeski gravity. These models introduce a non-minimal coupling between the scalar field and matter which preserves the second differential order of field equations. The most general beyond Horndeski non-minimal coupling is described by the following Lagrangian \cite{Kobayashi:2019hrl}:
\begin{align}\label{the_beyond_Horndeski_coupling}
  \mathcal{A}_\text{int} =  \int d^4 x \sqrt{-g}\left[ C(\phi, X)\, g_{\mu\nu}+ D(\phi, X) \, \pd_\mu\phi \, \pd_\nu\phi \right] T^{\mu\nu} .
\end{align}
Here $C$ and $D$ are arbitrary functions of the scalar field $\phi$ and the canonical kinetic term $X$.

It must be noted that both Horndeski and beyond Horndeski models are well motivated and widely studied \cite{Latosh:2018xai,Starobinsky:2016kua,Babichev:2013cya,Babichev:2016rlq,Mironov:2019mye,Volkova:2019jlj,Dyadina:2018ryl}. In particular, beyond Horndeski models originated from degenerate higher-order scalar-tensor (DHOST) theories \cite{Horndeski:2016bku,Motohashi:2016ftl,Motohashi:2017eya,Motohashi:2018pxg,Bekenstein:1992pj,Zumalacarregui:2013pma,Langlois:2017mxy,deRham:2016wji} and it was shown that only models \eqref{the_beyond_Horndeski_coupling} have no ghost modes in the spectrum of tensor perturbations propagating about a cosmological background \cite{Langlois:2017mxy,deRham:2016wji}. More detailed reviews of these models can be found elsewhere \cite{Kobayashi:2019hrl,Achour:2016rkg}.

Interactions taking place within beyond Horndeski (and other DHOST) theories become relevant in the strong field regime. In particular, they can affect cosmological evolution resulting in an existence of a late-time de Sitter attractor \cite{Crisostomi:2018bsp}, gradient instabilities \cite{Libanov:2016kfc,Creminelli:2016zwa}, existence of bouncing solutions \cite{Kolevatov:2017voe}. Moreover they can lead to null energy condition violations \cite{Kobayashi:2016xpl} and an existence of stable wormholes \cite{Mironov:2018uou,Franciolini:2018aad}.

Within modified gravity the usage of Horndeski and beyond Horndeski models is well-motivated, as these models, in some sense, describe the most general class of physically acceptable models. In this paper we argue that some beyond Horndeski interactions are generated dynamically within quantum theory. 

Quantum behavior of a gravity theory can be consistently described within effective field theory formalism \cite{Donoghue:1994dn,Burgess:2003jk,Barvinsky:1985an}. Such an approach to gravity is well studied for general relativity and it allows one to obtain some verifiable predictions \cite{BjerrumBohr:2002kt,Bjerrum-Bohr:2016hpa,Latosh:2020jyq}. We also should note that one should not neglect the whole manifold of alternative approaches to quantum features of scalar-tensor theories \cite{Barra:2019kda,Buchbinder:2019bcc,Steinwachs:2011zs,Heisenberg:2020cyi,Saltas:2016nkg,Barvinsky:2008ia,Barvinsky:2009fy,Barvinsky:2009ii}.

Recently it was found that a certain non-minimal interaction between the scalar field and gravity is dynamically induced at the one-loop level \cite{Latosh:2020jyq}. The effect is a complete analogy with the anomalous fermion dipole moment. In this paper we show that this non-minimal interaction is also responsible for a dynamical generation of a certain beyond Horndeski interaction. We also show that some other beyond Horndeski interactions can be generated at the loop level, although they will be strongly suppressed. 

This paper is organized as follows. In Section \ref{section_anomalous_John} we briefly discuss the effective field theory method, its implementation to gravity, and the mechanism responsible for the generation of a new non-minimal coupling between the scalar field and gravity. In Section \ref{section_anomalous_beyond_Horndeski} we show that the same mechanism generates certain beyond Horndeski interactions at the one-loop level. We discuss what interactions can be generated by such a mechanism and point to the fact that some of them are suppressed more then the others. We highlight the interaction which provides the leading contribution in the low energy regime. We conclude in Section \ref{section_conclusion} where we discuss the physical role of such interactions.

\section{Anomalous non-minimal coupling}\label{section_anomalous_John}

Effective field theory method originated within particle physics \cite{Georgi:1994qn} and recently was applied to gravity \cite{Donoghue:1994dn,Burgess:2003jk,Barvinsky:1985an}. It must be highlighted that a few different approaches are usually labeled as effective field theory. Perhaps, the effective potential technique, bases on a resummation of connected one-particle irreducible diagrams, is the most known one \cite{Buchbinder:1992rb}. In this paper we are intend to used a different technique which is based on a premise of a factorization of ultraviolet physics. In the context of gravity this means that low energy phenomena, for instance, Solar system physics, are not affected by the Planck scale fluctuations in a meaningful way. This allows one to set a factorization scale\footnote{In the previous papers \cite{Latosh:2020jyq,Arbuzov:2020pgp} is was called ``the renormalization scale''. Such a terminology is misleading and we will not use it hereafter.} $\mu$, which is naturally should lie below the Planck scale, and to construct a low energy theory as a momentum expansion. This construction factorizes out all phenomena lying above the factorization scale $\mu$ and provides a consistent quantum description of the low energy physics.

On the practical ground such an approach to effective field theory paradigm is implemented to gravity as follows. Firstly, one sets the factorization scale $\mu$ which lies below the Planck mass. Secondly, one defines the microscopic action $\mathcal{A}$ of a gravity theory. The action describes both gravitation and non-gravitational degrees of freedom. Finally, one performs a background field quantization \cite{DeWitt:1967yk,DeWitt:1967ub,DeWitt:1967uc,Abbott:1983zw}.

Background field quantization, in turn, is perform as follows. Firstly, one finds a suitable background metric $\overline{g}_{\mu\nu}$ which solves the classical field equations. Then, one introduces the full metric $g_{\mu\nu}$ which accounts both for the background contribution and for the perturbations propagating about it:
\begin{align}
  g_{\mu\nu} = \overline{g}_{\mu\nu} + \kappa \, h_{\mu\nu} .
\end{align}
Here $\kappa$ is a dimensional parameter (in our case it is related with the Newtonian constant $G_N$ as $\kappa^2 = 32 \pi G_N$) and $h_{\mu\nu}$ is a field with the canonical mass-dimension which describes small metric perturbations. Lastly, the action $\mathcal{A}$ is expanded in a series with respect to $h_{\mu\nu}$:
\begin{align}
  \begin{split}
    \mathcal{A}[g] = \mathcal{A}[\overline{g}] + \cfrac{\delta\mathcal{A}}{\delta g_{\mu\nu}}~ \kappa h_{\mu\nu} + \cfrac{1}{2} \, \cfrac{\delta\mathcal{A}}{\delta g_{\mu\nu} \delta g_{\alpha\beta}} ~\kappa^2 \, h_{\mu\nu} h_{\alpha\beta} + \cfrac{1}{3!} \,\cfrac{\delta\mathcal{A}}{\delta g_{\mu\nu} \delta g_{\alpha\beta} \delta g_{\rho\sigma}} ~ \kappa^3 \, h_{\mu\nu} + O(\kappa^4).
  \end{split}
\end{align}
After that the field $h_{\mu\nu}$ is quantized. In this expansion $\delta^2 A/ (\delta g_{\mu\nu} \, \delta g_{\alpha\beta} )$ terms describe propagation of a free graviton, while other terms describe gravitational field self-interaction and interaction with matter.

Such an approach to effective gravity is discussed in great details in papers \cite{Burgess:2003jk,Donoghue:1994dn}, so we will not discuss it further. We would only highlight two things. The first one is the gauge fixing. We use the de Donder gauge $\pd_\mu h^{\mu\nu} - \cfrac12\, \pd^\nu\, h=0$ and fix it with the following gauge-fixing term:
\begin{align}
  \mathcal{L}_\text{gf} = \left(\pd_\mu h^{\mu\nu} - \cfrac12\, \pd^\nu h\right)^2 .
\end{align}
The second one is that we use the flat metric $\eta_{\mu\nu}$ as the background.

An important milestone of effective gravity was found in paper \cite{Donoghue:1994dn} where it was shown that one can use this approach to evaluate quantum corrections to the classical Newton potential. Such calculations can be performed due to the structure of loop corrections. The power-law corrections of the Newton potential are given by non-analytical functions in the momentum space \cite{Donoghue:1994dn}:
\begin{align}\label{non-analytical_terms}
  \begin{split}
    &\int\cfrac{d^3 \vec k}{(2\pi)^3}\, \cfrac{1}{k} \, e^{-i \vec{k}\cdot\vec{r}} = \cfrac{1}{2\pi^2 r^2}\,, \\
    &\int\cfrac{d^3 \vec k}{(2\pi)^3} \, \ln(\vec k)^2 \, e^{-i \vec{k}\cdot\vec{r}} = -\cfrac{1}{2\pi^2 r^3}\,.
  \end{split}
\end{align}
Such non-analytical terms are typically generated by loop corrections. We shall clarify the relation between locality and analyticity. Formula \eqref{non-analytical_terms} shown that contributions that are non-analytic in the momentum space corresponds to long-range forces with an inverse power-law decay. Terms that are analytic in the momentum space, on the contrary, corresponds to interaction that contain derivatives, i.e. to interactions dependent on local gradients. In the low energy regime, where we perform the most part of terrestrial experiments, non-analytic terms dominate over local corrections which means that long-range forces dominate over local interactions. Most importantly, non-analytic terms are multiplied by finite coefficients and \textit{do not depend on the factorization scale}. To put it otherwise, these terms are protected from the high energy content of a given model and can be treated as a universal prediction.

A similar effect was found in paper \cite{Latosh:2020jyq}. In an analogy with \cite{Donoghue:1994dn} the paper addresses an effect that does not depend on the factorization scale. At the one-loop level the following non-minimal interaction between a scalar field and gravity is generated:
\begin{align}\label{anomalous_John_interaction}
  \mathcal{L} =  G^{\mu\nu} \, \nabla_\mu\phi\,\nabla_\nu\phi.
\end{align}
This interaction is generated by the following diagram:
\begin{align}\label{anomalous_John}
  \begin{gathered}
    \begin{fmffile}{D01}
      \begin{fmfgraph*}(50,50)
        \fmfleft{L}
        \fmfright{R1,R2}
        \fmf{dbl_wiggly,tension=2,label=$l$}{L,V}
        \fmf{dashes}{R1,D,V,U,R2}
        \fmf{dbl_wiggly,tension=.2,right=.3}{D,U}
        \fmflabel{$\mu\nu$}{L}
        \fmfdot{U,D,V}
        \fmffreeze
        \fmf{phantom}{R1,D}
        \fmf{phantom,label=$p$}{R1,D}
        \fmf{phantom,label=$q$}{R2,U}
      \end{fmfgraph*}
    \end{fmffile}
  \end{gathered}
  = i \kappa^3 \, l^2 \, C_{\mu\nu\alpha\beta} \, p^\alpha q^\beta.
\end{align}
The following definition of $C_{\mu\nu\alpha\beta}$ is used here and further:
\begin{align}
  C_{\mu\nu\alpha\beta} = \eta_{\mu\alpha}\eta_{\nu\beta} + \eta_{\mu\beta}\eta_{\nu\alpha} - \eta_{\mu\nu}\eta_{\alpha\beta}\,.
\end{align}
The amplitude contains a contribution free from ultraviolet divergences\footnote{The amplitude also contain terms related with external states renormalization \cite{Latosh:2020jyq}. We omit them, as they are irrelevant for the discussion.}. Infrared divergences of such an amplitude are regularized via soft graviton radiation. Therefore, in full analogy with the previous case, one finds a contribution that is independent on the high energy content of a model. Moreover, exactly the same phenomenon is well-known within particle physics where it is responsible for the generation of anomalous magnetic momenta.

This provides ground to claim that the non-minimal kinetic interaction \eqref{anomalous_John_interaction} is a universal prediction of quantum gravity which is independent on the high energy content of a model. It also should be noted that, unlike the other case \cite{Donoghue:1994dn}, such an interaction is local and strongly suppressed. Because of this one can establish some empirical constraints on its coupling \cite{Latosh:2020jyq}. 

\section{Anomalous beyond Horndeski couplings}\label{section_anomalous_beyond_Horndeski}

The effect founded in \cite{Latosh:2020jyq} shows the at the loop level gravity generates new interactions with a scalar field which are free from ultraviolet divergences. Naively, one may expect that this effect remains confined to the scalar sector of a theory, but this is not the case. If one introduces regular matter (standard model or beyond standard model particles minimally coupled to gravity), then loop effects are to introduce a non-minimal interaction between the scalar field and matter. This makes the main difference from the previously considered case \cite{Latosh:2020jyq}. The non-minimal interaction between a scalar field and gravity found in \cite{Latosh:2020jyq} does not remain confined to the scalar sector, but introduces new non-minimal interaction with the regular matter.

The same mechanism that generates the non-minimal kinetic coupling \eqref{anomalous_John_interaction} is responsible for a generation of a certain beyond Horndeski interaction. Let us consider a minimal scalar-tensor model which contains no non-minimal couplings
\begin{align}\label{the_model}
  \begin{split}
    \mathcal{A} = \int d^4 x \sqrt{-g} \Bigg[ -\cfrac{1}{16 \pi G}\, R +\cfrac12\,g^{\mu\nu}\, \nabla_\mu \phi \, \nabla_\nu\phi + \mathcal{L}_\text{matter} \left[\Psi,g_{\mu\nu}\right]~\Bigg].
  \end{split}
\end{align}
For the sake of clarity let us note that model \eqref{the_model} is a Horndeski model because it has second order field equations. However the results to be obtained will hold for an arbitrary scalar field as long as it admits the canonical kinetic term. We discuss the role of the Horndeski nature of the scalar field in the last section.
At the one-loop level the model generates the desired non-minimal kinetic coupling \cite{Latosh:2020jyq}. To show that beyond Horndeski interactions are generated in a way similar to the non-minimal coupling we find the specific one-loop amplitude that generates a beyond Horndeski interaction.

Beyond Horndeski interactions deal with a non-minimal interaction between the scalar field and matter \eqref{the_beyond_Horndeski_coupling}. Therefore one should study gravitational scattering of the scalar field on matter degrees of freedom. The simplest suitable amplitude corresponds to the following diagram
\begin{align}\label{the_main_amplitude}
  \begin{gathered}
    \begin{fmffile}{D02}
      \begin{fmfgraph}(50,50)
        \fmfleft{L1,L2}
        \fmfright{R1,R2}
        \fmf{dashes}{L1,L,L2}
        \fmf{dbl_plain}{R1,R,R2}
        \fmf{dbl_wiggly}{L,R}
        \fmfblob{10}{L}
        \fmfdot{R}
      \end{fmfgraph}
    \end{fmffile}
  \end{gathered}
\end{align}
Here we note matter degrees of freedom with a double plain line; the hatched circle, in turn, notes the vertex given in \eqref{anomalous_John}. We only consider tree-level interaction between the matter and gravity because of the two reasons. Firstly, such corrections lie beyond the scope of this paper and were extensively studied before \cite{Donoghue:1994dn,BjerrumBohr:2002kt,Akhundov:1996jd,Grisaru:1975ei}. Secondly, we are interested in the coupling between regular matter and the new scalar field. Corrections to a coupling between matter and gravity can only alter a coupling between the matter energy-tensor and gravity. Therefore they are irrelevant for our purpose. Finally, such corrections will introduce an additional factors $\kappa$, so the resulted amplitude receives additional suppression.

Because of this only corrections to the scalar-tensor sector are relevant. Such corrections were discussed in paper \cite{Latosh:2020jyq} and, as we have highlighted, it was shown that an anomalous interaction \eqref{anomalous_John} is generated. This exact anomalous contribution generates an anomalous interaction between matter and the scalar field:
\begin{align}\label{anomalous_bH}
  &
  \begin{gathered}
    \begin{fmffile}{D03}
      \begin{fmfgraph}(40,40)
	\fmfleft{L1,L2}
        \fmfright{R1,R2}
	\fmf{dashes}{L1,L,L2}
        \fmf{dbl_plain}{R1,R,R2}
	\fmf{dbl_wiggly}{L,R}
        \fmfblob{10}{L}
	\fmfdot{R}
      \end{fmfgraph}
    \end{fmffile}
  \end{gathered} 
  \begin{split}
    = i \kappa^3 l^2 C_{\mu\nu\alpha\beta} p^\alpha q^\beta \, i \,\cfrac{\frac12 \, C^{\mu\nu\rho\sigma}}{l^2}\, i \kappa \, C_{\rho\sigma\lambda\tau} \, T^{\lambda\tau}  =- 2 \, i\, \kappa^4 \,p^\alpha\,q^\beta C_{\alpha\beta\mu\nu}\,T^{\mu\nu} .
  \end{split}
\end{align}
Here $p$ and $q$ are momenta of the ingoing scalar particles. The expression \eqref{anomalous_bH} is local, so it corresponds to the following contact interaction:
\begin{align}
  \begin{gathered}
    \begin{fmffile}{D04}
      \begin{fmfgraph}(40,40)
        \fmfleft{L1,L2}
        \fmfright{R1,R2}
        \fmf{dashes}{L1,V,L2}
        \fmf{dbl_plain}{R1,V,R2}
        \fmfdot{V}
      \end{fmfgraph}
    \end{fmffile}
  \end{gathered}
  \begin{split}
    =& -2\, i \, \kappa^4 \, p^\alpha q^\beta \, C_{\alpha\beta\mu\nu}\, T^{\mu\nu} \leftrightarrow -4 \kappa^4 \, \pd_\mu \phi \, \pd_\nu \phi \, T^{\mu\nu} + 2 \kappa^4 (\pd\phi)^2 \, T.
  \end{split}
\end{align}
It can be seen clearly that this is a beyond Horndeski interaction \eqref{the_beyond_Horndeski_coupling} with the following parameters:
\begin{align}
  C(\phi,X) &= 2 \kappa^4 (\pd\phi)^2 ,& D(\phi, X) &= - 4 \kappa^4.
\end{align}

Let us, once again, return to out interpretation of this result. In the previous paper \cite{Latosh:2020jyq} it was shown that even in the minimal model gravity develops a non-minimal interaction between with a scalar field. Here we show that any regular matter minimally coupled to gravity does develop a non-minimal interaction with a scalar field due to the same effect. In spirit of \cite{Donoghue:1994dn} this interaction does not depend to the factorization scale and free from ultraviolet divergences. This interaction is also local which is important for the infrared limit of a theory, as the leading infrared contribution is provided by non-local operators \cite{Donoghue:1994dn,Vanhove:2021zel}. Because of this the interaction \eqref{anomalous_bH} cannot enter the effective action evaluated with one-particle irreducible diagrams \cite{Buchbinder:1992rb}, but it will contribute to scattering processes.

It should be noted that the amplitude \eqref{the_main_amplitude} is not the only amplitude that generates beyond Horndeski interactions. However, the other amplitudes experience stronger suppression. Namely, the following amplitude may also generate beyond Horndeski interaction:
\begin{align}
  \begin{gathered}
    \begin{fmffile}{D05}
      \begin{fmfgraph}(50,50)
        \fmfleft{L}
        \fmfright{R1,R2}
        \fmftop{T}
        \fmfbottom{B}
        \fmf{dbl_wiggly,tension=2}{L,V}
        \fmf{dbl_plain}{B,D,V,U,T}
        \fmf{dbl_wiggly}{D,VR,U}
        \fmf{dashes}{R1,VR,R2}
        \fmfdot{V,VR,D,U}
      \end{fmfgraph}
    \end{fmffile}
  \end{gathered}
\end{align}
Here double plain lines note matter degrees of freedom. This amplitude may very well generate an interaction between one graviton, two scalars, and two matter degrees of freedom that belong to the beyond Horndeski class:
\begin{align}
  h_{\mu\nu} ~ (\pd\phi)^2 ~ T^{\mu\nu} \leftrightarrow C(\phi,X)\,g_{\mu\nu} \, T^{\mu\nu}.
\end{align}
Firstly, this interaction is suppressed by the factor $\kappa^5$, so it does not provide the leading contribution and can be safely neglected for the time being. Secondly, simple dimensional considerations show that this amplitude develops $\log$-dependence of the factorization scale. Although such a dependence of the factorization scale is weak, it does not allow one to consider such a contribution independent from the high energy content of a theory.

The following amplitudes can be treated in a similar way:
\begin{align}
  \begin{gathered}
    \begin{fmffile}{D06}
      \begin{fmfgraph}(50,50)
        \fmfleft{L1,L2}
        \fmfright{R1,R2}
        \fmf{dashes}{L1,L,L2}
        \fmf{dbl_plain}{R1,R,R2}
        \fmf{dbl_wiggly,left=.5,tension=.3}{L,R,L}
        \fmfblob{10}{L}
	\fmfdot{R}
      \end{fmfgraph}
    \end{fmffile}
  \end{gathered} ,
  \begin{gathered}
    \begin{fmffile}{D07}
      \begin{fmfgraph}(50,50)
        \fmfleft{L1,L2}
        \fmfright{R1,R2}
        \fmf{dashes}{L1,L,L2}
        \fmf{dbl_plain}{R1,R,R2}
        \fmf{dbl_wiggly,left=.7,tension=.3}{L,R,L}
        \fmfblob{10}{L}
        \fmfdot{R}
        \fmffreeze
        \fmf{dbl_wiggly}{L,R}
      \end{fmfgraph}
    \end{fmffile}
  \end{gathered} , \cdots
\end{align}
They are suppressed, at very least, by factor $\kappa^{2N}$, where $N$ is the number of virtual gravitons. Therefore the amplitude \eqref{anomalous_bH}, considered above, provides the leading contribution.

This proves our original claim, so a certain beyond Horndeski interaction are generated at the one-loop level. The corresponding amplitude \eqref{anomalous_bH} is free from ultraviolet divergences and therefore are independent on the high energy content of a quantum gravity model. We discuss this result and its implications in the next Section.

\section{Discussion and conclusion}\label{section_conclusion}

We have shown that a certain beyond Horndeski interaction is universally generated within effective field theory. By the universality we mean only the fact that such interaction is free from ultraviolet divergences. One may argue that some heavy states may exist that can alter the strength of \eqref{anomalous_bH}. This may very well be the case, but discussion of such states lie beyond the scope of this paper. Our aim was to consider the minimal possible model and show that even without new heavy states beyond Horndeski interaction is generated at loop level. In that respect we should highlight that \eqref{anomalous_bH} is a universal prediction due to gravity which does not account for more sophisticated matter content.

In that sense, the universality is given by the fact that both analytical \cite{Latosh:2020jyq} and non-analytical \cite{Donoghue:1994dn} terms generated at the loop level are independent from the factorization scale $\mu$ which separating low and high energy physics. Beyond Horndeski interactions are generated in the same way. Namely, a specific one-loop amplitude \eqref{anomalous_bH} contains a part generating the non-minimal kinetic coupling \eqref{anomalous_John_interaction} which is independent from the factorization scale. Because of this the amplitude \eqref{anomalous_bH} generates a new non-minimal interaction between the scalar field and matter. This interaction belongs to beyond Horndeski class and it does not depend on the factorization scale.

This shows that certain beyond Horndeski interactions are generated in a way similar to anomalous magnetic momentum and to a certain non-minimal kinetic coupling \cite{Latosh:2020jyq}. This result is relevant within Horndeski theory as it shows the need to include such nonlinear interaction. They are generated dynamically, although suppressed, and their manifestations may very well become relevant \cite{Mironov:2020pqh,Volkova:2019jlj,Mironov:2019mye}.

Let us also address the role of Horndeski theories. As it was noted above, the studied model \eqref{the_model} does belong to Horndeski theories as it admits second order field equations. However the obtained results will also hold for an arbitrary scalar field as long as it admits the canonical kinetic term. We believe that the model provides an interesting example suggesting that there might be a deeper relation between Horndeski and beyond Horndeski theories yet to be found. As a trivial Horndeski model \eqref{the_model} generates finite beyond Horndeski interactions at the one-loop level it provides as explicit example of a relation between Horndeski and beyond Horndeski theories emerging at the one-loop level. In order to positively confirm such a relation a sophisticated analysis of one-loop structure of Horndeski models is required. It lies far beyond the scope of this paper and will be performed elsewhere.

Because the discussed beyond Horndeski interaction is generated at the loop level it experience an extremely strong suppression by factor $\kappa^4 \sim G^2$. An additional suppression is also due to the factor $\hbar$, which is essential for quantum effects. Because of this such interactions are local, so they can hardly be found in the low energy (large scale) experiments. The standard post-Newtonian (PN) expansion does not account for quantum effects, which are described by terms suppressed by $\hbar$. Therefore one should not expect that such an interaction influence the standard post-Newtonian physics. And even if one finds a way to negate $\hbar$ factor, the interaction still may only contribute to 2PN (post-Newtonian) order or higher \cite{Will:2014kxa,Will:1993ns}. Although such effects can be calculated \cite{Blanchet:2013haa,Damour:2009sm,Blanchet:1987wq,Avdeev:2020jqo,Dyadina:2019dsu,Dyadina:2018ryl}, it would be challenging to efficiently constraint them with the current empirical data. Because of this we believe that there is no simple way to constraint such interactions by data obtained in contemporary Universe.

We shall note that such a strong suppression of the discussed interactions is twofold. On the one hand, the strong suppression may help to avoid certain undesirable features of beyond Horndeski interaction which are gradient instabilities of superluminal motion \cite{Libanov:2016kfc,Creminelli:2016zwa,Mironov:2020pqh,Mironov:2019mye}. On the other hand, in the early universe, where a scalar field may very well be strong and may have big gradients, such interactions will experience a much weaker suppression. In particular they may provide a required energy condition violation to realize Galilean Genesis \cite{Creminelli:2010ba,Kobayashi:2016xpl,Libanov:2016kfc,Kolevatov:2017voe,Volkova:2019jlj}.

It shall be noted that in a generic Horndeski model which includes non-minimal couplings between a scalar field and gravity new beyond Horndeski interactions can experience much weaker suppression in a late universe. Although a complete analysis lie beyond the scope of this paper, still, some estimates can be obtained from dimensional considerations. For the sake of completeness let us firstly consider the minimal model \eqref{the_model} discussed in this paper. The magnitude of the generated beyond Horndeski interaction corresponds to the following mass scale:
\begin{align}
  \Lambda_\text{BH}^4= \left(\cfrac{1}{M_\text{Pl}^2}\right)^2\, T \, (\nabla\phi)^2\,.
\end{align}
In the cosmological regime $4$-derivatives $\nabla\phi$ are reduced to time derivatives $\dot\phi$. In a late universe the scalar field tends to have a linear profile such that $\dot\phi = m M_\text{Pl}$ where $M_\text{Pl}$ is the Planck mass and $m$ is a scalar field mass. The energy momentum tensor can be approximated by the contemporary matter density $T \sim \rho \sim 10^{-47} \text{ GeV}^4$. Therefore the characteristic beyond Horndeski scale reads:
\begin{align}
  \Lambda_\text{BH} \sim \sqrt{ \cfrac{m}{M_\text{Pl}}} \,10^{-11}\, \text{eV}\,.
\end{align}
In full agreement with the arguments given above such an interaction is strongly suppressed in the late-time universe.

In the case of a generic Horndeski model such estimates are altered. The reason is the fact that non-minimal Horndeski couplings between a scalar field and gravity can introduce two new independent energy scales defined by the Hubble scale $H$:
\begin{align}
  \Lambda_1^2 &= M_\text{Pl} \, H  & \Lambda_2^3 &= M_\text{Pl} \, H^2
\end{align}
The value of the contemporary Hubble parameter is $H_0 \sim 9\times 10^{-42}$ GeV, so mass scales $\Lambda_1$ and $\Lambda_2$ read:
\begin{align}
  \Lambda_1 \sim 10^{-2} \text{ eV }, ~\Lambda_2 \sim 9 \times 10^{-12} \text{ eV}\,.
\end{align}
The same linear profile $\dot\phi = M_\text{Pl} \, m$ can be used to estimate the magnitude of the derivative interactions. If the discussed beyond Horndeski interaction is suppressed by $\Lambda_1$ then its characteristic energy scale reads
\begin{align}
  \Lambda_\text{BH} \sim \sqrt{\cfrac{m}{M_\text{Pl}}} \, \times 10^{20} \text{ eV} \,.
\end{align}
If $\Lambda_2$ is used then the corresponding energy scale
\begin{align}
  \Lambda_\text{BH} \sim \sqrt{\cfrac{m}{M_\text{Pl}}} \, \times 10^{17} \text{ eV} \,.
\end{align}
In both cases the suppression is much weaker then in the simplest model. 

Finally, we would like to make a brief comment on an influence of loop corrections on the Horndeski structure. Horndeski theory can be viewed as a theory with the so-called Galileon symmetry \cite{Germani:2011bc,Deffayet:2013lga,Saltas:2016nkg}. This makes a theory more natural, as operators generating higher derivative contribution to field equations are strongly suppressed \cite{Pirtskhalava:2015nla}. In this paper we did not introduce Galileon symmetry of any kind, so it is natural to expect that the model develop higher-derivative operators at loop level. A simple dimensional considerations show that at the one-loop level one should expect that operators $\kappa^{2N} \square^2 \phi^{2N}$ will be generated. Therefore there is no a priory reason to believe that the theory preserves its second-order nature at the loop level. A more detailed discussion of this issue, however, lies far beyond the scope of this paper.

\section*{Acknowledgment}
The work was supported by the Foundation for the Advancement of Theoretical Physics and Mathematics “BASIS”.

\bibliographystyle{unsrturl}
\bibliography{BHdQE.bib}

\end{document}